\documentclass[12pt]{article}
\usepackage{amsmath}

\newcommand{\bm}{\begin{multiline}}
\newcommand{\beq}{\begin{equation}}
\newcommand{\eeq}{\end{equation}}
\newcommand{\beqs}{\begin{eqnarray}}
\newcommand{\eeqs}{\end{eqnarray}}

\numberwithin{equation}{section}

\begin{document}

\addtolength{\baselineskip}{1.5mm}

\thispagestyle{empty}

\begin{flushright}
hep-th/0505114
\end{flushright}
\hfill{}

\hfill{}

\hfill{}

\vspace{32pt}

\begin{center}
\textbf{\Large Higher dimensional Kaluza-Klein Monopoles} \\[0pt]

\vspace{48pt}
{\bf Robert Mann}\footnote{E-mail: 
{\tt rbmann@sciborg.uwaterloo.ca}}
{\bf and Cristian Stelea}\footnote{E-mail: 
{\tt cistelea@uwaterloo.ca}}

\vspace*{0.2cm}

{\it $^{1}$Perimeter Institute for Theoretical Physics}\\
{\it 31 Caroline St. N. Waterloo, Ontario N2L 2Y5 , Canada}\\[.5em]

{\it $^{1, 2}$Department of Physics, University of Waterloo}\\
{\it 200 University Avenue West, Waterloo, Ontario N2L 3G1, Canada}\\[.5em]
\end{center}

\vspace{48pt}

\begin{abstract}
\addtolength{\baselineskip}{1.2mm} It is well known that the Kaluza-Klein
monopole of Sorkin, Gross and Perry can be obtained from the Euclidean
Taub-NUT solution with an extra compact fifth spatial dimension via
Kaluza-Klein reduction. In this paper we consider Taub-NUT-like solutions of
the vacuum Einstein field equations, with or without cosmological constant,
in five dimensions and higher, and similarly perform Kaluza-Klein reductions
to obtain new magnetic KK brane solutions in higher dimensions, as well as further sphere reductions to
magnetic monopoles in four dimensions. In six dimensions we also employ spatial and timelike Hopf
dualities to untwist the circle fibration characteristic to these spaces and
obtain charged strings. A variation of these methods in ten dimensions leads
to a non-uniform electric string in five-dimensions.
\end{abstract}

\setcounter{footnote}{0}

\newpage

\section{Introduction}

Although magnetic monopole solutions have been found in vacuum Kaluza-Klein
theories since the $1980$s  \cite{sorkin,mp}, it remains an outstanding
problem to generalize these soliton solutions (here termed Kaluza-Klein
monopoles, or KK monopoles) to include a cosmological constant. Given
current theoretical interest in asymptotically (anti)-de Sitter spacetimes
and recent experimental results that indicate that the universe does indeed
possess a small positive cosmological constant, it is reasonable to pursue
such an objective.

One such attempt was recently made by Onemli and Tekin \cite{onemli}. They
concluded that there is no five-dimensional static Kaluza-Klein monopole
with cosmological constant. The metric ansatz they employed was tailored to
describe a static `Kaluza-Klein' monopole in an $AdS_{2}\times AdS_{3}$
background. In the limit in which the cosmological constant $\lambda $ tends
to zero, the `KK-AdS monopole' should reduce to the `KK monopole' in flat
space. Moreover, if the monopole charge tends to zero then
the `KK-AdS monopole' should reduce smoothly to the $AdS_{2}\times AdS_{3}$
background. If we relax the requirement that the monopole solution be static
it is easy to construct a time-dependent five-dimensional soliton that has
all the desired properties. Two such solutions in five dimensions have been
obtained in \cite{onemli}.

In the present paper we consider other possible extensions of Kaluza-Klein
monopole solutions that admit a cosmological constant. The essential
ingredient in the original Kaluza-Klein monopole construction is a Euclidean section
of a Taub-NUT-like space; the `trick' employed in \cite{sorkin,mp} to obtain
the monopole solution is to lift this Euclidean section up to
five-dimensions by adding a flat time coordinate and then to dimensionally
reduce along the `Euclidean time' direction from the Euclidean Taub-NUT section. However, in a presence of the cosmological constant it is not possible to use the above technique without introducing an explicit time dependence in the metric. Therefore, in order to obtain cosmological four-dimensional
magnetic monopole solutions our strategy is to consider directly in
five-dimensions the new cosmological Taub-NUT-like solutions\footnote{It is also worth mentioning that, in a different context \cite{d2}, some unexpected results were obtained for asymptotically AdS Taub-NUT spacetimes.},  recently obtained in \cite{csrm,Page1} and perform a Kaluza-Klein compactification along the fifth dimension. The new feature of these solutions is that the
four-dimensional dilaton acquires a potential term as an effect of the
cosmological constant. However their asymptotics are not very appealing
physically since they are not asymptotically flat or $(A)dS$ in the Einstein frame. Their metric description
simplifies when considered in the string frame: for our explicit examples
the four-dimensional metric in the string frame is very similar to the $AdS$
form in the $(r, t)$ sector, except for a deficit of solid angle in the
angular sector. Another interesting feature of the above constructions is that in five
dimensions and in some of the higher dimensional examples the nut charge and
the cosmological constant are intimately related by a constraint equation
imposed by the equations of motion. This constraint makes it impossible to
consider situations in which either the cosmological constant or the nut
charge go to zero\footnote{See \cite{rick} for a way to evade this constraint in some special cases.}.
From this perspective the above monopole solutions are qualitatively
distinct from their predecessors.

In higher than five dimensions we have more choices: we can consider solutions that
are Ricci flat with different nut parameters or we can consider Taub-NUT
like spaces that are constructed as circle fibrations over base spaces that
have non-trivial topology. We also perform Kaluza-Klein (KK) reductions of
the above solutions down to four dimensions, obtaining new magnetic monopole
solutions. More specifically, in six and seven dimensions we have considered
non-singular Ricci-flat solutions for which one can use the KK trick to
obtain similar KK magnetic brane solutions for which the background spaces
are Ricci flat Bohm spaces of the form $S^{p}\times S^{q}$ and generically
have conical singularities. We considered their further reduction down to
four dimensions on Riemannian spaces of constant curvature and specifically
considered such reductions on spheres. In contrast with the KK procedure to
untwist the $U(1)$-fibration, we have considered in six dimensions another
method that is known to untwist the circle fibration, namely Hopf duality
in string theory. We extended these duality rules to the case of a timelike
Hopf-duality of the truncated six-dimensional Type II theories and applied
them to generate charged string solutions in six-dimensions. By performing
sphere reductions we obtained the corresponding four-dimensional solutions. In general, the presence of the cosmological constant in the higher dimensional theory induces a scalar potential for the Kaluza-Klein scalar fields. If the isometry generated by the Killing vector $\frac{\partial }{\partial z}$, which is associated with the circle direction on which we perform the reduction has fixed points, then the dilaton, which describes the radius of that extra-dimension, will diverge at the fixed point sets and the $D$-dimensional metric will be singular at those points. In certain cases we find that the dilaton field also diverges at infinity. Respectively this means that, physically, the space-time decompactifies near the KK-brane and
at infinity; the higher-dimensional theory should be used when describing such objects in these regions.  

The organization of this paper is as follows. We begin in section \ref{GPSmonopole} by reviewing how the flat KK monopole can be obtained from the four dimensional Taub-nut solution. We also briefly discuss the KK monopole obtained by using the Euclidian Taub-bolt soliton. We next present in section \ref{newmonopole} the new metric ansatz which is a solution of vacuum Einstein's
equations with cosmological constant in five dimensions and we perform a
Kaluza-Klein reduction to obtain a new four-dimensional monopole solution.
In the following sections we consider similar monopole solutions in higher
dimensions and we also perform Kaluza-Klein sphere reductions to four
dimensions. In six dimensions we apply spatial and timelike Hopf-dualities
to generate new solutions. Section \ref{conclusion} concludes with some
comments, and in appendices, for convenience, the KK reduction formulae and
the $T$-duality rules which relate the truncated six-dimensional Type II
theories are summarized.

\section{The $GPS$ magnetic monopole}

\label{GPSmonopole}

We begin by reviewing the original magnetic monopole solution in $4$ dimensions that
arises as a Kaluza-Klein compactification of a $5$ dimensional vacuum metric \cite%
{sorkin,mp}. The essential ingredient used in the monopole construction is a 
$4$-dimensional version of the Taub-NUT solution, with Euclidean signature.
The monopole solution is constructed as follows.

Start with the Euclidean form of the Taub-NUT solution \cite{Misner}: 
\begin{equation*}
ds^{2}=F_{E}(r)(d\chi -2n\cos \theta d\varphi
)^{2}+F_{E}^{-1}(r)dr^{2}+(r^{2}-n^{2})d\Omega ^{2}
\end{equation*}%
where 
\begin{equation*}
F_{E}(r)=\frac{r^{2}-2mr+n^{2}}{r^{2}-n^{2}}
\end{equation*}%
In general, the $U(1)$ isometry generated by the Killing vector $\frac{%
\partial }{\partial \chi }$ (that corresponds to the coordinate $\chi $ that
parameterizes the fibre $S^{1}$) can have a zero-dimensional fixed point set
(referred to as a `nut' solution) or a two-dimensional fixed point set
(correspondingly referred to as a `bolt' solution). The regularity of the
Euclidean Taub-nut solution requires that the period of $\chi $ be $\beta =8\pi n$
(to ensure removal of the Dirac-Misner string singularity), $F_{E}(r=n)=0$
(to ensure that the fixed point of the Killing vector $\frac{\partial }{%
\partial \chi }$ is zero-dimensional) and also $\beta F_{E}^{\prime
}(r=n)=4\pi $ in order to avoid the presence of the conical singularities at 
$r=n$. With these conditions we obtain $m=n$, yielding 
\begin{equation*}
F_{E}(r)=\frac{r-n}{r+n}
\end{equation*}%
Taking now the product of this Euclidean space-time with the real line, we
obtain the following $5$-dimensional Ricci flat metric: 
\begin{equation*}
ds^{2}=-dt^{2}+F_{E}(r)(d\chi -2n\cos \theta d\varphi
)^{2}+F_{E}^{-1}(r)dr^{2}+(r^{2}-n^{2})d\Omega ^{2}
\end{equation*}%
which solves the $5$-dimensional vacuum Einstein equations.

If we perform now a Kaluza-Klein reduction along the coordinate $\chi $
(which is periodic with period $8\pi n$) we obtain the following $4$%
-dimensional fields (with $\alpha =\frac{1}{2\sqrt{3}}$) \cite{sorkin,mp} 
\begin{eqnarray}
ds^{2} &=&-F_{E}^{\frac{1}{2}}dt^{2}+F_{E}^{-\frac{1}{2}}(r)dr^{2}+F_{E}^{%
\frac{1}{2}}(r^{2}-n^{2})d\Omega ^{2}  \notag \\
\mathcal{A} &=&-2n\cos \theta d\varphi,~~~~~~~ e^{\frac{\phi }{\sqrt{3}}}
=F_{E}^{-\frac{1}{2}}  \label{sorkinmonopole}
\end{eqnarray}%
It is clear now that the metric is asymptotically flat. The above solution
describes a magnetic monopole and its properties have been discussed in
detail in \cite{sorkin,mp}.

There are a few extensions of the above construction that we can consider.
The obvious one to explore is the Taub-bolt solution in four-dimensions
instead of the nut solution. In this case the Killing vector $\frac{\partial 
}{\partial \chi }$ has a two-dimensional fixed point set in the $4$-dimensional Euclidean sector. The regularity of the solution is then
ensured by the following conditions  \cite{Page,Mann,Chamblin}: $%
F(r=r_{b})=0$ and $\frac{4\pi }{F^{\prime }(r_{b})}=\frac{8\pi n}{k}$ where $%
k$ is an integer while the period of $\chi $ is now given by $\beta =\frac{%
8\pi n}{k}$, i.e. we identify $k$ points on the circle described by $\chi $.

It is easy to see that the above conditions are satisfied for $r_{b}=\frac{2n%
}{k}$ and $m=m_{p}=\frac{n(4+k^{2})}{4k}$. We must demand that $r\geq r_{b}>n$,
so that the fixed point set of $\frac{\partial }{\partial \chi }$ is not
zero-dimensional; this in turn avoids the curvature singularity at $r=n$ and
forces $k=1$. Then the period of the coordinate $\chi $ is $8\pi n$ and for
the bolt solution we obtain \cite{Page}: 
\begin{equation}
F_{E}(r)=\frac{\left( r-2n\right) \left( r-\frac{1}{2}n\right) }{r^{2}-n^{2}}\label{FEbolt}
\end{equation}
As in the case of the nut solution, we take the product with the real line
and obtain a metric in five-dimensions that is a solution of the vacuum
Einstein field equations. Performing the Kaluza-Klein compactification along
the $\chi $ direction yields (\ref{sorkinmonopole}), where now $F_E(r)$ is given by (\ref{FEbolt}). The five-dimensional metric is regular everywhere for $r\geq 2n$. However,
the four-dimensional solution obtained by Kaluza-Klein
reduction, while asymptotically flat, is now singular at the location of the
bolt $r=2n$ where the dilaton field diverges as expected.

The physical interpretation of this solution was recently clarified by Liang
and Teo \cite{Edward1} (see also \cite{Chamblin1}). It corresponds to a pair
of coincident extremal dilatonic black holes with opposite magnetic charges.
To see this we can use as a seed in the KK procedure the Euclidean rotating
version of the bolt solution \cite{Demiansky,Gibbons:nf}. We add a timelike
flat direction in order to lift the solution to five dimensions, after which
we reduce down to four dimensions. When $n\neq 0$ it has been shown in \cite{Edward1} that the above solution describes a pair of extremal dilatonic
black holes carrying opposite but unbalanced magnetic charges and separated
by a distance $2a$, $a$ being the rotation parameter, which in this case
serves as a measure of the proper distance between the black holes. In the
limit $a\rightarrow 0$ we obtain the the solution (\ref{sorkinmonopole}) and (\ref{FEbolt}) which
corresponds then to a pair of coincident monopoles that carry opposite
unbalanced magnetic charges. The total magnetic charge of the system is $n$.

Since the above dihole has unbalanced charges it is not possible to introduce a background magnetic field to stabilize the system \cite{Edward1}. Consequently the solution is unstable and is expected to decay to a pure nut solution (the GPS soliton in this case) with total charge $n$.

\section{Taub-Nut-dS/AdS spacetimes in $5$ dimensions}

\label{newmonopole}

In four dimensions the usual Taub-NUT construction corresponds to a $U(1)$%
-fibration over a two-dimensional Einstein space used as the base space.
Usually taken to be the sphere $S^{2}$, it can also be the torus $T^{2}$ or
the hyperboloid $H^{2}$. In five dimensions the corresponding base space is
three dimensional and consequently the above construction is not straightforward.
However, we can construct the five-dimensional Taub-NUT space as a sphere
fibration over a factor space $S^{2}$ of the base space \cite{csrm,Page1}.
The spacetimes that we obtain are not trivial in the sense that there is now
a constraint on the possible values of the nut charge and the cosmological
constant. Specifically, we cannot simultaneously set the nut charge and/or
the cosmological constant to zero: there is no smooth limit in which one can
obtain five dimensional Minkowski space in this way.

The ansatz that we shall use in the construction of these spaces is the
following 
\begin{equation}
ds^{2}=-F(r)(dt-2n\cos \theta d\varphi
)^{2}+F^{-1}(r)dr^{2}+(r^{2}+n^{2})(d\theta ^{2}+\sin ^{2}\theta d\varphi
^{2})+r^{2}dz^{2}  \label{TNB5d}
\end{equation}%
The above metric will be a solution of the Einstein field equations with
positive cosmological constant $\lambda =\frac{6}{l^{2}}$ provided 
\begin{equation*}
F(r)=\frac{4ml^{2}-r^{4}-2n^{2}r^{2}}{l^{2}(r^{2}+n^{2})}
\end{equation*}%
where the field equations impose the constraint $4n^{2}=l^{2}$. Notice that
for large values of $r$ the function $F(r)$ takes negative values and $r$
becomes effectively a timelike coordinate. We can write the solution
directly in Euclidean signature (analytically continuing $t\rightarrow i\chi 
$ and $n\rightarrow in$): 
\begin{equation}
ds^{2}=F_{E}(r)(d\chi -2n\cos \theta d\varphi
)^{2}+F_{E}^{-1}(r)dr^{2}+(r^{2}-n^{2})(d\theta ^{2}+\sin ^{2}\theta
d\varphi ^{2})+r^{2}dz^{2}
\end{equation}%
where 
\begin{equation}
F_{E}(r)=\frac{4r^{4}-2l^{2}r^{2}+16ml^{2}}{l^{2}(4r^{2}-l^{2})}
\label{FE5dold}
\end{equation}%
and we restrict the values of $r$ to be greater than the largest root of $%
F_{E}$ (if it has any) or than $l/2$, in order to preserve the metric
signature. Since the analytic continuation of $n$ forces the continuation $%
l\rightarrow il$ for consistency with the initial constraint equation we
obtain an Einstein metric with negative cosmological constant $\lambda =-%
\frac{6}{l^{2}}$.

In order to remove the usual Misner string singularity in the metric, we have
to assume that the coordinate $\chi $ is periodic with some period $\beta $.
Notice that for $r=n$ the fixed point of the Killing vector $\frac{\partial}{%
\partial \chi }$ is one dimensional (i.e. less than its maximal possible
co-dimension) and we have a Taub-nut solution. However, for $r=r_{b}$, where $%
r_{b}>n$ is the largest root of $F_{E}(r)$, the fixed point set is
three-dimensional (its maximal possible value); we shall refer to such
solutions as Taub-bolt solutions.

As before, in order to have a regular nut solution we must ensure the same
conditions on the periodicity of $\chi $ and the relationship $\beta
F_{E}^{\prime }(r=n)=4\pi $. This leads to $m=%
\frac{l^{2}}{64}$; it is precisely for this value of the parameter $m$ that
the above solution becomes the Euclidean AdS spacetime in five-dimensions.

The bolt solutions likewise have the regularity conditions given in the
previous section: the period of $\chi $ is $\beta =\frac{8\pi n}{k}=\frac{%
4\pi }{F_{E}^{\prime }(r_{b})}$, where k is an integer and $r=r_{b}>n$.
These yield $r_{b}=\frac{kn}{2}$ and 
\begin{equation}
m=m_{b}=\frac{k^{2}l^{2}(k^{2}-8)}{1024}  \label{mbolt5d}
\end{equation}%
Setting $k\geq 3$ to ensure that $r_{b}>n$ (thereby also avoiding the
curvature singularity at $r=n$), we obtain the following family of bolt
solutions, indexed by the integer $k$: 
\begin{equation}
ds^{2}=F_{E}(r)(d\chi -l\cos \theta d\phi )^{2}+F_{E}^{-1}(r)dr^{2}+\bigg[%
r^{2}-\left( \frac{l}{2}\right) ^{2}\bigg](d\theta ^{2}+\sin ^{2}\theta
d\phi ^{2})+r^{2}dz^{2}
\end{equation}%
where 
\begin{equation}
F_{E}(r)=\frac{256r^{4}-128l^{2}r^{2}-k^{2}l^{4}(k^{2}-8)}{%
64l^{2}(4r^{2}-l^{2})}  \label{FE5d}
\end{equation}

However, for our purposes we need a five-dimensional solution that has
Lorentzian signature. To obtain such a solution we shall analytically
continue the coordinate $z\rightarrow it$. We then obtain the following
five-dimensional space-time: 
\begin{equation}
ds^{2}=F_{E}(r)(d\chi -l\cos \theta d\phi )^{2}+F_{E}^{-1}(r)dr^{2}+\bigg[%
r^{2}-\left( \frac{l}{2}\right) ^{2}\bigg](d\theta ^{2}+\sin ^{2}\theta
d\varphi ^{2})-r^{2}dt^{2}  \label{5dKKseed}
\end{equation}%
which is a solution of the vacuum Einstein field equations with a negative
cosmological constant. Only when $k\geq 3$ (ensuring that $r_{b}=\frac{%
3l}{4}>\frac{l}{2}$) and $m=m_{b}$\ from (\ref{mbolt5d}) is inserted into the metric
function (\ref{FE5dold}) will the solution (\ref{5dKKseed}) have no
curvature singularities; otherwise there is a curvature singularity at $r=\frac{l}{2}$.

\subsection{Monopole solutions in $4$ dimensions}

We are now ready to generate the magnetic monopole solutions in four
dimensions using the same procedure as in the $GPS$ case. Since the
Taub-nut/bolt solution (\ref{TNB5d}) is non-singular if we periodically identify the
coordinate $\chi $ with period $\beta $, we can perform a Kaluza-Klein
reduction along this direction. The Kaluza-Klein ansatz that we shall use is
given in Appendix A, with $\alpha =\frac{1}{2\sqrt{3}}$.

Since our initial space-time is a solution of the Einstein field equations
with non-zero cosmological constant, the field content in four dimensions
will be given by the metric tensor $g_{\mu \nu }$, a magnetic one-form
potential $\mathcal{A}$ and a scalar field $\phi $ with a non-trivial scalar
potential $V(\phi )$. We obtain 
\begin{eqnarray}
ds^{2} &=&-r^{2}F_{E}^{\frac{1}{2}}dt^{2}+F_{E}^{-\frac{1}{2}%
}(r)dr^{2}+F_{E}^{\frac{1}{2}}\left(r^{2}-\frac{l^{2}}{4}\right)d\Omega ^{2}
\notag \\
\mathcal{A} &=&-2n\cos \theta d\varphi,~~~~~~~ e^{\frac{\phi }{\sqrt{3}}}
=F_{E}^{-\frac{1}{2}}  \label{gen4}
\end{eqnarray}
where 
\begin{equation*}
F_{E}(r)=\frac{4r^{4}-2l^{2}r^{2}+16ml^{2}}{l^{2}(4r^{2}-l^{2})}
\end{equation*}

This solution differs from the Kaluza-Klein $GPS$ solution in terms of both
the metric coefficients and by the fact that the scalar field $\phi $ has
a potential of the exponential type%
\begin{equation*}
V(\phi )=-\frac{8}{l^{2}}e^{-\frac{\phi }{\sqrt{3}}}
\end{equation*}%
indicating that our Kaluza-Klein dimensional reduction yields a massive
scalar field.

To study the properties of the above $4$-dimensional spaces, let us consider
first the Taub-nut solution, which we have seen corresponds to the AdS solution
in five-dimensions. For $m=\frac{l^{2}}{64}$ we obtain (with $n=\frac{l}{2}$%
): 
\begin{equation*}
F_{E}(r)=\frac{4r^{2}-l^{2}}{4l^{2}}
\end{equation*}
and the five-dimensional metric becomes: 
\begin{equation*}
ds^{2}=-(1+\frac{R^{2}}{l^{2}})dt^{2}+\frac{1}{1+\frac{R^{2}}{l^{2}}}dR^{2}+%
\frac{R^{2}}{4}\bigg[\frac{1}{l^2}(d\chi -\cos \theta d\varphi )^{2}+d\theta
^{2}+\sin ^{2}\theta d\varphi ^{2}\bigg]
\end{equation*}
where we have defined $R=4r^2-l^2$ and rescaled the coordinates $t$ and $%
\chi $.

After Kaluza-Klein compactification we obtain the following four-dimensional
fields: 
\begin{eqnarray}
ds^{2} &=&-\frac{R}{2l}(1+\frac{R^{2}}{l^{2}})dt^{2}+\frac{RdR^{2}}{%
2l\left(1+\frac{R^{2}}{l^{2}}\right)}+\frac{R^{3}}{8l}d\Omega ^{2}  \notag \\
\mathcal{A} &=&-l\cos \theta d\varphi,~~~~~~~ e^{-\frac{\phi }{\sqrt{3}}} =%
\frac{R}{2l}
\end{eqnarray}
The four-dimensional metric has a curvature singularity at $R=0$, where the
scalar field also diverges. Its asymptotic structure is given by 
\begin{equation*}
ds^{2}=-\frac{\tilde{R}^{2}}{l^{2}}dt^{2}+\frac{4}{9}\left( \frac{l}{\tilde{R%
}}\right) ^{\frac{4}{3}}d\tilde{R}^{2}+\tilde{R}^{2}d\Omega ^{2}
\end{equation*}
where we have rescaled $t$ and defined $\tilde{R}$ by $\tilde{R}^{2}=\frac{%
R^{3}}{8l}$ and we can see that it is not asymptotically flat. However, one
can check by computing some of the curvature scalars (like $R_{abcd}R^{abcd}$%
) that the above asymptotic metric is well-behaved at infinity and that it
has a curvature singularity at $\tilde{R}=0$. Notice however that at both $%
R=0$ and $R\rightarrow\infty$ the dilaton is blowing up and the extra
dimension opens up, which means that the physical description is effectively
five-dimensional.

For the bolt solution we must take $k\geq 3$ and $m=m_{p}$, for which 
\begin{equation*}
F_{E}(r)=\frac{256r^{4}-128l^{2}r^{2}-k^{2}l^{4}(k^{2}-8)}{64l^{2}(4r^{2}-l^{2})}
\end{equation*}%
The compactified four-dimensional solution is given again by (\ref{gen4})
and its asymptotic structure is the same as the one obtained from the
five-dimensional $AdS$ spacetime. Again the four-dimensional metric will
have a curvature singularity at the bolt while the dilaton diverges at $%
r=r_{b}$ and at infinity.

Next, let us notice that in the string frame the situation changes as
follows. The five-dimensional nut solution reduces to the following metric: 
\begin{equation*}
ds^{2}=-(1+\frac{R^{2}}{l^{2}})dt^{2}+\frac{1}{1+\frac{R^{2}}{l^{2}}}dR^{2}+%
\frac{R^{2}}{4}\bigg[d\theta ^{2}+\sin ^{2}\theta d\varphi ^{2}\bigg]
\end{equation*}%
While this metric resembles a four-dimensional AdS metric with cosmological
constant $\lambda =-\frac{3}{l^{2}}$ in fact there is a deficit of solid
angle as the area of the 2-sphere is $\pi R^{2}$ instead of $4\pi R^{2}$.
This behavior is characteristic of a global monopole \cite{vilenkin}. The above metric has a curvature singularity at the origin
(the location of the monopole) and the dilaton field diverges both at origin
and at infinity. The magnetic charge is computed using the formula \cite{sparks}: 
\begin{equation*}
\frac{1}{4\pi }\int_{S^{2}}\mathcal{F}=l
\end{equation*}%

Note that if we reduce directly the five-dimensional metric (\ref{TNB5d}),
which is time-dependent (outside the cosmological horizon the coordinate $r$
is timelike) we obtain a time-dependent four-dimensional magnetic monopole
solution.

One characteristic feature of the above constructions is that the nut charge
and the cosmological constant are intimately related by a constraint
equation imposed by the equations of motion. This constraint makes it
impossible for us to consider the cases in which either the cosmological
constant or the nut charge go to zero. From this perspective the above
monopole solutions are qualitatively distinct from their predecessors.

\section{Higher dimensional magnetic monopoles}

\label{higherdim}

We now consider some of the higher dimensional Taub-NUT spaces constructed
recently \cite{csrm,Page1}. In the Taub-NUT ansatz the idea is to construct
such spaces as radial extensions of $U(1)$-fibrations over an even
dimensional base endowed with an Einstein-K$\ddot{a}$hler metric. In a $%
(2k+2)$-dimensional Taub-NUT space the base factor over which one constructs
the circle fibration can have at most dimension $2k$. However, it is not
necessary to construct the circle fibration over the whole base space; in
general one can consider factorizations of the base of the form $B=M\times Y$
in which $M$ is endowed with an Einstein-K$\ddot{a}$hler metric while $Y$ is
an Einstein space with the metric $g_{Y}$. In these cases one can consider
the $U(1)$-fibration only over the factor space $M$ of the base and take
then a warped product with the manifold $Y$. The ansatz is then given by 
\begin{equation}
F^{-1}(r)dr^{2}+(r^{2}+N^{2})g_{M}+r^{2}g_{Y}-F(r)(dt+A)^{2}
\end{equation}
We now consider particular cases of this ansatz.

\subsection{Six dimensional metrics}

In six dimensions the base space is four-dimensional and we can use products of the form $%
M_{1}\times M_{2}$ of two-dimensional Einstein spaces or we can use $CP^{2}$
as a four-dimensional base space over which to construct the circle
fibrations. If we use products of two dimensional Einstein spaces then we
can consider all the cases in which $M_{i}$, $i=1,2$ can be a sphere $S^{2}$%
, a torus $T^{2}$ or a hyperboloid $H^{2}$. The circle fibration can be
constructed in these cases over the whole base space $M_{1}\times M_{2}$ or
just over one factor space $M_{i}$.

We shall consider first the case in which $M_{1}=M_{2}=S^{2}$ and assume
that the $U(1)$ fibration is constructed over the whole base space $%
S^{2}\times S^{2}$. Then the corresponding six-dimensional Taub-NUT solution
is given by \cite{csrm} 
\begin{eqnarray}
ds^{2} &=&-F(r)(dt-2n_{1}\cos \theta _{1}d\varphi _{1}-2n_{2}\cos \theta
_{2}d\varphi _{2})^{2}+F^{-1}(r)dr^{2}  \notag \\
&&+(r^{2}+n_{1}^{2})(d\theta _{1}^{2}+\sin ^{2}\theta _{1}d\varphi
_{1}^{2})+(r^{2}+n_{2}^{2})(d\theta _{2}^{2}+\sin ^{2}\theta _{2}d\varphi
_{2}^{2})
\end{eqnarray}%
where 
\begin{eqnarray}
F(r) &=&\frac{%
3r^{6}+(l^{2}+5n_{2}^{2}+10n_{1}^{2})r^{4}+3(n_{2}^{2}l^{2}+10n_{1}^{2}n_{2}^{2}+n_{1}^{2}l^{2}+5n_{1}^{4})r^{2}%
}{3(r^{2}+n_{1}^{2})(r^{2}+n_{2}^{2})l^{2}}  \notag \\
&&+\frac{6ml^{2}r-3n_{1}^{2}n_{2}^{2}(l^{2}+5n_{1}^{2})}{%
3(r^{2}+n_{1}^{2})(r^{2}+n_{2}^{2})l^{2}}  \label{6nn}
\end{eqnarray}%
Here the above metric is a solution of vacuum Einstein field equations with
cosmological constant ($\lambda =-\frac{10}{l^{2}}$) if and only if $%
(n_{1}^{2}-n_{2}^{2})\lambda =0$. Consequently we see that differing values
for $n_{1}$ and $n_{2}$ are possible only if the cosmological constant
vanishes. For $n_{1}=n_{2}=n$ the above solution reduces to the
six-dimensional solution found in \cite{awad,bais}. In what follows we shall look
at the case of two different nut charges, that is we set the cosmological
constant to zero.

Let us consider the Euclidean section, obtained by the following analytic
continuations $t\rightarrow i\chi $ and $n_{j}\rightarrow in_{j}$ where $%
j=1,2$: 
\begin{eqnarray}
ds^{2} &=&F_{E}(r)(d\chi -2n_{1}\cos \theta _{1}d\varphi _{1}-2n_{2}\cos
\theta _{2}d\varphi _{2})^{2}+F_{E}^{-1}(r)dr^{2}  \notag \\
&&+(r^{2}-n_{1}^{2})(d\theta _{1}^{2}+\sin ^{2}\theta _{1}d\varphi
_{1}^{2})+(r^{2}-n_{2}^{2})(d\theta _{2}^{2}+\sin ^{2}\theta _{2}d\varphi
_{2}^{2})
\end{eqnarray}%
where 
\begin{equation}
F_{E}(r)=\frac{r^{4}-3(n_{1}^{2}+n_{2}^{2})r^{2}+6mr-3n_{1}^{2}n_{2}^{2}}{%
3(r^{2}-n_{1}^{2})(r^{2}-n_{2}^{2})}
\end{equation}

This metric is a solution of the vacuum Einstein field equations without
cosmological constant, for any values of the parameters $n_{1}$ and $n_{2}$.
In order to analyze the possible singularities we have to consider two
cases, depending on the values of the nut charges $n_{1}$ and $n_{2}$. Let
us assume first that $n_{1}=n_{2}=n$. Then the Taub-nut solutions correspond to
the fixed-point sets of $\frac{\partial }{\partial \chi }$ at $r=n$. The
mass parameter $m=\frac{4n^{3}}{3}$, while the periodicity of the coordinate 
$\chi $ is $12\pi n$. For the bolt solution we impose the periodicity of the
coordinate $\chi $ to be $\frac{12\pi n}{k}$. The fixed-point set is
four-dimensional and located at $r=r_{p}=\frac{3n}{k}>n$ , with the mass
parameter given by $m=m_{p}=\frac{n^{3}(k^{4}+18k^{2}-27)}{6k^{3}}$. Hence
we must consider only the values $k=1,2$, which in turn avoids the curvature
singularity at $r=n$.

If the two nut charges $n_{1}$ and $n_{2}$ differ, we set $n_{1}>n_{2}$
without loss of generality. In this case in the Euclidean section the radius 
$r$ cannot be smaller than $n_{1}$ or the signature of the spacetime will
change. The Taub-nut solution in this case corresponds to a two-dimensional
fixed-point set located at $r=n_{1}$. There is still a curvature singularity
located at $r=n_{1}$, removed by setting the periodicity of the coordinate $%
\chi $ to be $8\pi n_{1}$, while the value of the mass parameter must be $%
m=m_{p}=\frac{n_{1}^{3}+3n_{1}n_{2}^{2}}{3}$.

The bolt solution corresponds to a four-dimensional fixed-point set located
at $r=r_{b}=\frac{2n_{1}}{k}$, for which the periodicity of the coordinate $%
\chi $ is given by $\frac{8\pi n_{1}}{k}$ and the value of the mass
parameter is $m=m_{p}=\frac{n_{1}(12n_{2}^{2}-4n_{1}^{2})}{12}$. In order to
avoid the curvature singularity at $r=n_{1}$ we must choose $k=1$ so that $%
r>r_{b}=2n_{1}$.

In the following we consider the Taub-nut solution for which $n_{1}>n_{2}$. The
periodicity of the coordinate $\chi $ is taken to be $8\pi n_{1}$ while the
value of the mass parameter is fixed to be $m=m_{p}=\frac{%
n_{1}^{3}+3n_{1}n_{2}^{2}}{3}$. For these values the six-dimensional metric
is nonsingular at $r=n_{1}$.

Employing the usual Kaluza-Klein procedure we obtain a six-dimensional
magnetic monopole: we add a flat time direction to obtain a
seven-dimensional solution of the vacuum Einstein field equations and after
that perform a Kaluza-Klein compactification along the coordinate $\chi $
using the metric ansatz: 
\begin{equation*}
ds_{7}^{2}=e^{\frac{\phi }{\sqrt{10}}}ds_{6}^{2}+e^{-\frac{4\phi }{\sqrt{10}}%
}(d\chi -A_{(1)})^{2}
\end{equation*}

It is easy to check that we obtain the following six-dimensional fields 
\begin{eqnarray}
ds_{6}^{2} &=&-F_{E}^{\frac{1}{4}}dt^{2}+F_{E}^{-\frac{3}{4}}dr^{2}+F_{E}^{%
\frac{1}{4}}\left[ (r^{2}-n_{1}^{2})(d\theta _{1}^{2}+\sin ^{2}\theta
_{1}d\varphi _{1}^{2})+(r^{2}-n_{2}^{2})(d\theta _{2}^{2}+\sin ^{2}\theta
_{2}d\varphi _{2}^{2})\right]  \notag \\
A_{(1)} &=&-2n_{1}\cos \theta _{1}d\varphi _{1}-2n_{2}\cos \theta
_{2}d\varphi _{2},~~~~~~~ e^{-\frac{\phi }{\sqrt{10}}} =F_{E}^{\frac{1}{4}}
\end{eqnarray}
where we now restrict $r\geq n_{1}>n_{2}$ and 
\begin{equation*}
F_{E}(r)=\frac{r^{3}+n_{1}r^{2}-(2n_{1}^{2}+3n_{2}^{2})r+3n_{1}n_{2}^{2}}{%
3(r+n_{1})(r^{2}-n_{2}^{2})}
\end{equation*}

One can check that the above six-dimensional monopole solution has a
curvature singularity located at $r=n_{1}$. It is interesting to note that
the asymptotic structure of this solution, after rescaling the coordinates $%
t\rightarrow 3^{1/4}T$ and $r\rightarrow 3^{-3/8}R$ is given by 
\begin{equation*}
ds_{asymp}^{2}=-dT^{2}+dR^{2}+\frac{1}{3}R^{2}(d\theta _{1}^{2}+\sin
^{2}\theta _{1}d\varphi _{1}^{2})+\frac{1}{3}R^{2}(d\theta _{2}^{2}+\sin
^{2}\theta _{2}d\varphi _{2}^{2})
\end{equation*}%
The area of each 2-sphere is not $4\pi R^{2}$ but instead $\frac{4\pi R^{2}%
}{3}$: each has a deficit solid angle of $\frac{8\pi }{3}$ steradians.
Furthermore, the above asymptotic form of the metric is Ricci flat, and can
be obtained from our solution by setting $n=0$. Therefore we conclude that
the background for our monopole is a Ricci flat Bohm metric constructed as a
cone over $S^{2}\times S^{2}$ \cite{Bohm,Gibbons}.

The corresponding six-dimensional Lagrangian obtained after Kaluza-Klein
reduction is given by 
\begin{equation*}
\mathcal{L}_{6}=eR-\frac{1}{2}e(\partial \phi )^{2}-\frac{1}{4}ee^{-\frac{5}{%
\sqrt{10}}\phi }F_{(2)}^{2}
\end{equation*}
where $F_{(2)}=dA_{(1)}=2n_{1}\Omega _{1}+2n_{2}\Omega _{2}$ and we have
denoted by $\Omega _{i}$ the volume form $\sin \theta _{i}d\theta _{i}\wedge
d\varphi _{i}$ of the sphere $M_{i}$, $i=1,2 $.

In the following we shall perform a Kaluza-Klein reduction on $M_{2}$. The
general sphere reduction formulae have been presented in \cite{bremer}. The
metric ansatz that we have to use in the dimensional reduction from six to
four dimensions is given by: 
\begin{equation*}
ds_{6}^{2}=e^{\frac{\varphi }{\sqrt{2}}}ds_{4}^{2}+e^{-\frac{\varphi }{\sqrt{%
2}}}(d\theta _{2}^{2}+\sin ^{2}\theta _{2}d\varphi _{2}^{2})  \label{6s}
\end{equation*}
The dimensionally-reduced Lagrangian will take now the form 
\begin{equation*}
\mathcal{L}_{4}=eR-\frac{1}{2}e(\partial \varphi )^{2}-\frac{1}{2}e(\partial
\phi )^{2}-\frac{1}{4}ee^{-\frac{\varphi }{\sqrt{2}}-\frac{5}{\sqrt{10}}\phi
}F^{2}+ee^{\sqrt{2}\varphi }R_{2}-e2n_{2}^{2}e^{\frac{3}{\sqrt{2}}\varphi -%
\frac{5}{\sqrt{10}}\phi }
\end{equation*}
where $F=dA=d(-2n_{1}\cos \theta _{1}d\phi _{1})$ and $R_{2}=4$ is the Ricci
scalar of the sphere $M_{2}$. The full solution in four dimensions will be
given by: 
\begin{eqnarray}
ds_{4}^{2} &=&-F_{E}^{\frac{1}{2}}(r^{2}-n_{2}^{2})dt^{2}+F_{E}^{-\frac{1}{2}%
}(r^{2}-n_{2}^{2})dr^{2}+F_{E}^{\frac{1}{2}%
}(r^{2}-n_{1}^{2})(r^{2}-n_{2}^{2})(d\theta _{1}^{2}+\sin ^{2}\theta
_{1}d\varphi _{1}^{2})  \notag \\
F &=&dA=2n_{1}\sin \theta _{1}d\theta _{1}\wedge d\varphi _{1} ,~~~~~ e^{-%
\frac{\phi }{\sqrt{10}}} =F_{E}^{\frac{1}{4}},~~~~~ e^{-\frac{\varphi }{%
\sqrt{2}}} =F_{E}^{\frac{1}{4}}(r^{2}-n_{2}^{2})
\end{eqnarray}

The asymptotic form of the above four-dimensional monopole metric is given
by: 
\begin{equation}
ds_{asymp}^{2}\sim -Rdt^{2}+dR^{2}+\frac{4R^{2}}{3}(d\theta _{1}^{2}+\sin
^{2}\theta d\varphi _{1}^{2})  \label{6d1}
\end{equation}
(after defining $r^{2}=\frac{2R}{3^{\frac{1}{4}}}$ and rescaling the time
coordinate $t$) and we can see that the spacetime is not asymptotically
flat. Moreover the metric has infinite redshift at the origin\footnote{A similar BPS monopole solution with an infinite redshift at the origin but with a deficit of solid angle has been obtained in \cite{hartnoll}.}, which is also the location of a curvature singularity. It takes an infinite time for
a photon to reach infinity and, indeed, the $(r,t)$-sector is asymptotically
flat; however, while the metric it is singularity-free at infinity the
scalar field $\varphi$ diverges there. It is interesting to note that the
asymptotic form has a surfeit of solid angle, as the area of a sphere of
radius $R$ is not $4\pi R^{2}$ but $\frac{16\pi r^{2}}{3}$. Asymptotically conical metrics are reminiscent of global monopoles \cite{vilenkin}. The magnetic charge is computed to be $2n_1$.

The second case to discuss in six dimensions is a generalization of the
five-dimensional solution presented in the previous section. The metric
ansatz is as follows: 
\begin{eqnarray}
ds^{2} &=&-F(r)(dt-2n\cos \theta _{1}d\phi _{1})^{2}+F^{-1}(r)dr^{2}
+(r^{2}+n^{2})(d\theta _{1}^{2}+\sin ^{2}\theta _{1}d\phi _{1}^{2})+\alpha
r^{2}(d\theta _{2}^{2}+d\phi _{2}^{2})  \notag \\
F(r)&=&\frac{-3r^{5}+(l^{2}-10n^{2})r^{3}+3n^{2}(l^{2}-5n^{2})r+6ml^{2}}{%
3rl^{2}(r^{2}+n^{2})}
\end{eqnarray}%
and the vacuum Einstein field equations with cosmological constant are
satisfied if and only if $\alpha (2-\lambda n^{2})=0$. Since $\alpha $
cannot be zero we must restrict the values of $n$ and $\lambda =\frac{10}{%
l^{2}}$ such that $\lambda n^{2}=2$, forcing a positive cosmological
constant. The Euclidean section is obtained by taking the analytic
continuation $t\rightarrow i\chi $ and $n\rightarrow in$ and $l\rightarrow il
$ with $l=\sqrt{5}n$. Notice that in this case the Taub-nut solution corresponds
to a two-dimensional fixed-point set of the vector field $\frac{\partial }{%
\partial \chi }$ located at $r=n$. The periodicity of the $\chi $ coordinate
is in this case equal to $8\pi n$ and the value of the mass parameter is
fixed to $m_{b}=\frac{n^{3}}{15}$. For this value of the mass parameter the
solution is regular at the nut location. The bolt spacetime has a
four-dimensional fixed-point set of $\frac{\partial }{\partial \chi }$
located at $r_{b}=\frac{kn}{2}$ and the value of the mass parameter is $%
m_{b}=\frac{k^{3}n^{3}(20-3k^{2})}{960}$ where the periodicity of the
coordinate $\chi $ is $\frac{8\pi n}{k}$, where $k$ is an integer. To ensure
that $r_{b}>n$ we have to take $k>3$ ; in this way the curvature singularity
at $r=n$ is avoided as well. We next perform another analytic continuation
of one of the coordinates on $T^{2}$ (say $\theta _{2}\rightarrow it$) and
then two Kaluza-Klein reductions along the $\chi $ and $\phi _{2}$
directions down to four dimensions. We obtain the final solution: 
\begin{eqnarray}
ds^{2} &=&-r^{3}F_{E}^{\frac{1}{2}}dt^{2}+F_{E}^{-\frac{1}{2}%
}(r)rdr^{2}+F_{E}^{\frac{1}{2}}r\left( r^{2}-\frac{l^{2}}{5}\right) d\Omega
^{2}  \notag \\
\mathcal{A} &=&-2n\cos \theta _{1}d\phi _{1},~~~~~ e^{\frac{-3\varphi _{1}}{%
\sqrt{6}}} =r^{2},~~~~~ e^{\frac{-\varphi _{2}}{\sqrt{3}}} =r^{\frac{1}{3}%
}F_{E}^{-\frac{1}{2}}  \label{6T2}
\end{eqnarray}%
where 
\begin{equation*}
F_{E}(r)=\frac{15r^{5}-5l^{2}r^{3}+30ml^{2}}{3l^{2}r(5r^{2}-l^{2})}
\end{equation*}%
which is a solution of the equations of motion derived from the following
Lagrangean: 
\begin{equation*}
\mathcal{L}_{4}=eR-\frac{1}{2}e(\partial \varphi _{1})^{2}-\frac{1}{2}%
e(\partial \varphi _{2})^{2}-\frac{1}{4}ee^{-\sqrt{3}\varphi _{2}}\mathcal{F}%
^{2}+2ee^{\frac{\varphi _{1}}{\sqrt{6}}+\frac{\varphi _{2}}{\sqrt{3}}}\lambda
\end{equation*}%
with $\lambda =-\frac{10}{l^{2}}$ and $\mathcal{F}=d\mathcal{A}$. The
asymptotic form of the metric is: 
\begin{equation*}
ds_{4}^{2}=-R^{2}dt^{2}+\frac{\sqrt{l}}{4R}dR^{2}+R^{2}d\Omega ^{2}
\end{equation*}%
after rescaling $R^{2}=\frac{r^{4}}{l}$. It is clearly not asymptotically
flat, it has infinite redshift at origin, which is also the location of a
curvature singularity and as expected the dilaton fields diverge at
infinity. However, notice that while $\varphi _{2}$ diverges at the root of $%
F_{E}$, $\varphi _{1}$ is finite there. The magnetic charge is found to be $2n$.

\subsubsection{Hopf reductions in six dimensions}

It is well-known that odd-dimensional spheres $S^{2n+1}$ may be regarded as
circle bundles over $CP^{n}$ and one can use the so-called Hopf duality (a
T-duality along the $U(1)$-fibre) to generate new solutions \cite{Duff1,Duff2,Cvetic} by untwisting $S^{2n+1}$ to $CP^{n}\times S^{1}$. The six-dimensional case is particularly interesting for us since it has been shown in \cite{Duff2} that it is possible to make consistent truncations of the
maximal Type II supergravity theories to a bosonic sector which exhibits an $%
O(2,2)$ global symmetry with the $T$-duality transformation taking a simple
form. The theories at hand are the toroidal reductions of Type IIA,
respectively Type IIB ten-dimensional supergravities while the reduction
ansatz for the fields is that the six-dimensional fields that are retained
are precisely the ten-dimensional ones, with the spacetime indices
restricted to run over the six-dimensional range only. The two truncated
theories in $D=6$ are then related by a T-duality transformation upon
reduction to $D=5$. The explicit mappings of the fields have been given in %
\cite{Duff2} and we follow their notational conventions. For convenience we
also provide the derivation of the $T$-duality rules in Appendix B.

Let us start with the solution given in (\ref{6nn}) in which we set $\lambda
=0$. We shall perform first the analytic continuations \cite{d1} $t\rightarrow iz$, $%
n_{1}\rightarrow in_{1}$ and subsequently $\varphi _{2}\rightarrow it$: 
\begin{eqnarray}
ds^{2} &=&\tilde{F}(r)(dz-2n_{1}\cos \theta _{1}d\varphi _{1}-2n_{2}\cos
\theta _{2}dt)^{2}+\tilde{F}^{-1}(r)dr^{2}  \notag \\
&&+(r^{2}-n_{1}^{2})(d\theta _{1}^{2}+\sin ^{2}\theta _{1}d\varphi
_{1}^{2})+(r^{2}+n_{2}^{2})(d\theta _{2}^{2}-\sin ^{2}\theta _{2}dt^{2})
\label{6nin}
\end{eqnarray}%
where 
\begin{equation*}
\tilde{F}(r)=\frac{r^{4}-3(n_{1}^{2}-n_{2}^{2})r^{2}+6mr+3n_{1}^{2}n_{2}^{2}%
}{3(r^{2}-n_{1}^{2})(r^{2}+n_{2}^{2})}
\end{equation*}%
Considering the above metric as a solution of the pure gravity sector of the
truncated Type IIA theory we can now perform a Hopf-duality along the
spacelike $z$-direction to obtain a solution of six-dimensional Type IIB
theory: 
\begin{eqnarray}
ds_{6B} &=&\tilde{F}(r)^{-\frac{1}{2}}dz^{2}+\tilde{F}(r)^{-\frac{1}{2}%
}dr^{2}+\tilde{F}(r)^{\frac{1}{2}}(r^{2}-n_{1}^{2})d\Omega _{1}^{2}+\tilde{F}%
(r)^{\frac{1}{2}}(r^{2}+n_{2}^{2})(d\theta _{2}^{2}-\sin ^{2}\theta
_{2}dt^{2})  \notag \\
e^{2\phi _{1}} &=&e^{2\phi _{2}}=\tilde{F}(r),~~~~~~~ A_{(2)}^{NS}
=-2n_{1}\cos \theta _{1}d\varphi _{2}\wedge dz+2n_{2}\cos \theta
_{2}dt\wedge dz
\end{eqnarray}%
We can also make the analytic continuations $t\rightarrow i\varphi _{2}$ and 
$n_{1}\rightarrow in_{1}$ to obtain the solution: 
\begin{eqnarray}
ds_{6B} &=&-F(r)^{-\frac{1}{2}}dt^{2}+F(r)^{-\frac{1}{2}}dr^{2}+F(r)^{\frac{1%
}{2}}(r^{2}+n_{1}^{2})d\Omega _{1}^{2}+F(r)^{\frac{1}{2}}(r^{2}+n_{2}^{2})d%
\Omega _{2}^{2}  \notag \\
e^{2\phi _{1}} &=&e^{2\phi _{2}}=F(r),~~~~~~~ A_{(2)}^{NS} =2n_{1}\cos
\theta _{1}d\varphi _{2}\wedge dt-2n_{2}\cos \theta _{2}d\varphi _{2}\wedge
dt  \label{6Bfinal}
\end{eqnarray}

Were we to consider (\ref{6nin}) as a solution of the pure gravity sector of
Type IIB theory, then after performing the spacelike Hopf dualisation we
would obtain as an intermediate step 
\begin{eqnarray}
ds_{6A} &=&\tilde{F}(r)^{-\frac{1}{2}}dz^{2}+\tilde{F}(r)^{-\frac{1}{2}%
}dr^{2}+\bar{F}(r)^{\frac{1}{2}}(r^{2}-n_{1}^{2})d\Omega _{1}^{2}+\bar{F}%
(r)^{\frac{1}{2}}(r^{2}+n_{2}^{2})(d\theta _{2}^{2}-\sin ^{2}\theta
_{2}dt^{2})  \notag \\
e^{2\phi _{1}} &=&e^{2\phi _{2}}=\bar{F}(r),~~~~~~~ A_{(2)} =-2n_{1}\cos
\theta _{1}d\varphi _{2}\wedge dz+2n_{2}\cos \theta _{2}dt\wedge dz
\end{eqnarray}%
Performing the analytic continuations we recover (\ref{6Bfinal}) except that
we have now to replace $A_{(2)}^{NS}$ with $A_{(2)}$.

It is interesting to note that we can perform directly a timelike Hopf
reduction in six-dimensions\footnote{In which case we do not need to perform any analytical continuations.}. In this case if we start with a solution of
Type IIA theory by performing the timelike T-duality we obtain a solution of
the appropriate truncation of Type IIB$^{\ast }$ theory \cite{Hull}. If we
start instead with a solution of Type IIB theory and perform a timelike Hopf
duality we end up with a solution of an appropriate truncation of Type IIA$%
^{\ast }$ theory. The details of these reductions are gathered in the
Appendix B.

As an example we shall perform a Hopf duality starting from Type IIA theory.
Consider (\ref{6nn}) as a solution of the pure gravity sector of the
truncated six-dimensional Type IIA theory. Then the final solution of Type
IIB$^*$ will be given by: 
\begin{eqnarray}
ds_{6B^*}&=&-F(r)^{-\frac{1}{2}}dt^2+F(r)^{-\frac{1}{2}}dr^2+F(r)^{\frac{1}{2}%
}(r^2+n_1^2)d\Omega_1^2+F(r)^{\frac{1}{2}}(r^2+n_2^2)d\Omega_2^2  \notag \\
e^{2\phi_1}&=&e^{2\phi_2}=F(r),~~~~~~~
A_{(2)}^{NS}=2n_1\cos\theta_1d\varphi_1\wedge dt-2n_2\cos\theta_2
d\varphi_2\wedge dt
\end{eqnarray}
where now 
\begin{equation*}
F(r)=\frac{r^{4}-3(n_{1}^{2}+n_{2}^{2})r^{2}+6mr-3n_{1}^{2}n_{2}^{2}}{%
3(r^{2}+n_{1}^{2})(r^{2}+n_{2}^{2})}
\end{equation*}
If we start with (\ref{6nn}) as a solution of Type IIB then peforming a
timelike Hopf dualisation we obtain a similar solution of Type IIA$^*$ for
which: 
\begin{eqnarray}
A_{(2)}&=&2n_1\cos\theta_1d\varphi_1\wedge dt-2n_2\cos\theta_2
d\varphi_2\wedge dt
\end{eqnarray}

As we can see, some of the solutions obtained for Type IIA (respectively
IIB) and Type IIA$^{\ast }$ (respectively IIB$^{\ast }$) are identical after
we perform appropriate analytic continuations\footnote{
Which keep the metric and the fields real.}. This is to be expected once we
notice that they are solutions of the $NSNS$-sector only, which is the same
for both theories (their actions would differ only by the sign of the
kinetic terms of the $RR$-fields).

As an application of these solutions let us set for convenience $n_{2}=0$ in
(\ref{6Bfinal}) and perform a sphere reduction using the ansatz (\ref{6s})
down to a four-dimensional solution: 
\begin{eqnarray}
ds_{4B} &=&-r^{2}dt^{2}+r^{2}dr^{2}+F(r)r^{2}(r^{2}+n_{1}^{2})d\Omega
_{1}^{2}  \notag \\
e^{-\sqrt{2}\phi } &=&r^{4}F(r),~~~~~ e^{2\phi _{1}} =e^{2\phi
_{2}}=F(r),~~~~~ A_{(2)}^{NS} =2n_{1}\cos \theta _{1}d\varphi _{2}\wedge dt
\label{4Bfinal}
\end{eqnarray}%
which is a solution of the equations of motion derived from the following
Lagrangian: 
\begin{equation*}
\mathcal{L}_{4B}=eR-\frac{1}{2}e(\partial \varphi )^{2}-\frac{1}{2}%
e(\partial \phi _{1})^{2}-\frac{1}{2}e(\partial \phi _{2})^{2}-\frac{1}{4}%
ee^{-\frac{\varphi }{\sqrt{2}}-\varphi _{1}-\varphi
_{2}}(F_{(3)}^{NS})^{2}+ee^{\sqrt{2}\varphi }R_{2}
\end{equation*}%
The asymptotic form of the metric (\ref{4Bfinal}) is (after defining $R=%
\frac{r^{2}}{2}$ and rescaling $t$) 
\begin{equation*}
ds_{4B}\sim -Rdt^{2}+dR^{2}+\frac{4}{3}R^{2}d\Omega _{1}^{2}
\end{equation*}%
Amusingly, the asymptotic form of the metric is the same with (\ref{6d1}). The magnetic charge is found to be $2n$ and notice that there is an excess of solid angle as the area of the asymptotic sphere is $\frac{16\pi R^{2}}{3}$ instead of the expected $4\pi R^{2}$.

\subsection{Monopoles in $D\geq 7$ dimensions}

Similarly, we can construct Kaluza-Klein monopoles in seven and higher
dimensions. For example, in seven dimensions the base space is
five-dimensional and can be factorized in the form $B=M\times Y$, where $M$
is an even dimensional space endowed with an Einstein-K$\ddot{a}$hler metric
and $Y$ is a Riemannian Einstein space.

Let us consider the case in which $M=S^{2}$ while $Y$ can be a sphere $S^{3}$%
, a torus $T^{3}$ or a hyperboloid $H^{3}$. The solution is \cite{csrm} 
\begin{eqnarray}
ds^{2} &=&-F(r)(dt^{2}+2n\cos \theta d\varphi
)^{2}+F^{-1}(r)dr^{2}+(r^{2}+n^{2})(d\theta ^{2}+\sin ^{2}\theta d\varphi
^{2}) +\beta r^{2}g_{Y}  \notag
\end{eqnarray}%
where $g_{Y}$ is the metric on the unit-sphere $S^{3}$, torus $T^{3}$ or
hyperboloid $H^{3}$: 
\begin{equation}
F(r)=\frac{%
4r^{6}+(l^{2}+12n^{2})r^{4}+2n^{2}(l^{2}+6n^{2})r^{2}+4ml^{2}+n^{4}(l^{2}+6n^{2})%
}{4l^{2}r^{2}(r^{2}+n^{2})}
\end{equation}%
The cosmological constant is $\lambda =-\frac{15}{l^{2}}$ and the parameters 
$\beta $, $n$ and $\lambda $ are constrained via the relation $\beta
(5-2\lambda n^{2})=10k$, where $k=1,0,-1$ for $S^{3}$, $T^{3}$ and $H^{3}$
respectively. We must have $\beta >0$, which in turn imposes a joint
constraint on $\lambda n^{2}$ that can be satisfied in various ways
depending on the value of $k$. Since we are interested in a Ricci flat
solution we shall consider $\lambda =0$ and also $k=1$, in which case $%
Y=S^{3}$ and $\beta =2$. The Euclidean section of this solution, which is
obtained by analytic continuation of the coordinate $t\rightarrow i\chi $
and of the parameter $n\rightarrow in$ is given by: 
\begin{eqnarray}
ds^{2} &=&F_{E}(r)(d\chi ^{2}+2n\cos \theta d\varphi
)^{2}+F_{E}^{-1}(r)dr^{2}+(r^{2}-n^{2})(d\theta ^{2}+\sin ^{2}\theta
d\varphi ^{2})+2r^{2}g_{Y}  \notag
\end{eqnarray}%
where 
\begin{equation}
F(r)=\frac{r^{4}-2n^{2}r^{2}+n^{4}+4m}{4r^{2}(r^{2}-n^{2})}
\end{equation}%
The metric has a scalar curvature singularity located at $r=n$ as can be
checked by computing for instance the Kretschman scalar. However, if we take 
$m=0$ then the metric is well-behaved at $r=n$. Furthermore, if $r<n$ the
signature of the metric is unphysical; therefore we can restrict ourselves to
the interval $r\geq n$ for which the solution is non-singular. In order to
obtain the magnetic brane in seven-dimensions we employ the usual
procedure: add a flat timelike direction and compactify the Ricci-flat
eight-dimensional metric using the ansatz 
\begin{equation*}
ds_{8}^{2}=e^{\frac{\phi }{\sqrt{15}}}ds_{7}^{2}+e^{-\frac{5\phi }{\sqrt{15}}%
}(d\chi -A_{(1)})^{2}
\end{equation*}%
We obtain the following seven-dimensional fields: 
\begin{eqnarray}
ds_{7}^{2} &=&-F_{E}^{\frac{1}{5}}dt^{2}+F_{E}^{-\frac{4}{5}}dr^{2}+F_{E}^{%
\frac{1}{5}}\left( (r^{2}-n^{2})(d\theta ^{2}+\sin ^{2}\theta d\varphi
^{2})+2r^{2}d\Omega _{3}^{2}\right)  \notag \\
A_{(1)} &=&-2n\cos \theta d\varphi,~~~~~~~ e^{-\frac{\phi }{\sqrt{15}}}
=F_{E}^{\frac{1}{5}}
\end{eqnarray}%
where now $r\geq n$ and 
\begin{equation*}
F_{E}(r)=\frac{r^{2}-n^{2}}{4r^{2}}
\end{equation*}%
which are a solution of the equations of motion derived from the following
Lagrangian: 
\begin{equation*}
\mathcal{L}_{7}=eR-\frac{1}{2}e(\partial \phi )^{2}-\frac{1}{4}ee^{-\frac{6}{%
\sqrt{15}}\phi }F_{(2)}^{2}
\end{equation*}%
The above seven-dimensional metric has a curvature singularity at $r=n$. Its
asymptotic structure, after we rescale the coordinates $t\rightarrow
4^{1/10}T$ and $r\rightarrow 4^{-2/5}R$, is given by 
\begin{equation*}
ds_{asymp}^{2}=-dT^{2}+dR^{2}+\frac{R^{2}}{4}(d\theta ^{2}+\sin ^{2}\theta
d\varphi ^{2})+2^{3/5}R^{2}d\Omega _{3}^{2}
\end{equation*}%
This space has a deficit of solid angle corresponding to the sphere $S^{2}$
while the factor $S^{3}$ has a surfeit of solid angle. 

Let us perform now a further dimensional reduction of the above
seven-dimensional solution on the three-sphere $S^{3}$. The metric ansatz
that we can use in the dimensional reduction from $7$ to $4$ dimensions is
given by: 
\begin{equation*}
ds_{7}^{2}=e^{\frac{3\varphi }{2\sqrt{15}}}ds_{4}^{2}+e^{-\frac{\varphi }{%
\sqrt{15}}}d\Omega _{3}^{2}
\end{equation*}
The four-dimensional fields will be given by 
\begin{eqnarray}
ds_{4}^{2} &=&2^{\frac{3}{2}}r^{3}\left( -F_{E}^{\frac{1}{2}}dt^{2}+F_{E}^{%
\frac{1}{2}}dr^{2}+F_{E}^{\frac{1}{2}}(r^{2}-n^{2})(d\theta ^{2}+\sin
^{2}\theta d\varphi ^{2})\right)  \notag \\
A_{(1)} &=&-2n\cos \theta d\varphi,~~~~~ e^{-\frac{\phi }{\sqrt{15}}}
=F_{E}^{\frac{1}{5}},~~~~~ e^{-\frac{\varphi }{\sqrt{15}}} =2r^{2}F_{E}^{%
\frac{1}{5}}
\end{eqnarray}
and they are a solution of the equations of motion derived from the
following dimensionally reduced Lagrangian: 
\begin{equation*}
\mathcal{L}_{4}=eR-\frac{1}{2}e(\partial \varphi )^{2}-\frac{1}{2}%
e(\partial\phi )^{2}-\frac{1}{4}ee^{-\frac{3\varphi }{2\sqrt{15}}-\frac{6}{%
\sqrt{15}}\phi }F_{(2)}^{2}+ee^{\frac{5}{\sqrt{15}}\varphi }R_{3}
\end{equation*}
where $R_{3}=6$ is the curvature scalar of the unit sphere $S^{3}$.

One can check that the above four-dimensional solution has a scalar
curvature singularity at $r=n$. Its asymptotics are given by: 
\begin{equation*}
ds^{2}=R^{3}\bigg[-dT^{2}+dR^{2}+R^{2}(d\theta ^{2}+\sin ^{2}\theta d\varphi
^{2})\bigg]
\end{equation*}
after we make the rescaling $R=\sqrt{2}r$ and $T=t/2$. Consequently the
spacetime that we obtain is conformally flat and singularity free at
infinity. The magnetic charge is found to be $2n$.

Generalization to more than seven dimensions is straightforward. For
instance, in eight dimensions we can consider a circle fibration over the
product $S^{2}\times S^{2}\times S^{2}$ and associate for each sphere factor a nut
charge $n_{i}$, $i=1..3$. We obtain the following metric 
\begin{eqnarray}
ds^{2} &=&-F(r)(dt+\mathcal{A})^{2}+\frac{dr^{2}}{F(r)}%
+\sum_{i=1,3}(r^{2}+n_{i}^{2})d\Omega _{i}^{2}  \notag \\
F(r) &=&\frac{%
3r^{6}+5(n_{1}^{2}+n_{2}^{2}+n_{3}^{2})r^{4}+15(n_{1}^{2}n_{2}^{2}+n_{2}^{2}n_{3}^{2}+n_{1}^{2}n_{3}^{2})r^{2}-15mr-15n_{1}^{2}n_{2}^{2}n_{3}^{2}%
}{15(r^{2}+n_{1}^{2})(r^{2}+n_{2}^{2})(r^{2}+n_{3}^{3})}  \notag \\
\mathcal{A} &=&2n_{1}\cos \theta _{1}d\phi _{1}+2n_{2}\cos \theta _{2}d\phi
_{2}+2n_{3}\cos \theta _{3}d\phi _{3}
\end{eqnarray}
that is a natural generalization of the eight-dimensional solution presented
in \cite{bais}. Our metric contain three different nut parameters. Taking
its Euclidean section (obtained by analytically continuing $t\rightarrow i\chi $, $%
n_{j}\rightarrow in_{j}$ for $j=1..3$) and adding a trivial time direction
we can then perform Kaluza-Klein reductions along the $\chi $ direction and
also along two of the spheres to obtain a final four-dimensional
monopole solution.

We can also consider different factorizations of the base space in terms of
products of Einstein-K$\ddot{a}$hler manifolds with various lower
dimensional Einstein spaces (as in \cite{csrm}).

Another interesting solution can be found in eleven dimensions by using the
ansatz: 
\begin{equation*}
ds^{2}=-F(r)(dt+2n\cos \theta d\phi )^{2}+\frac{dr^{2}}{F(r)}%
+(r^{2}+n^{2})(d\theta ^{2}+\sin ^{2}\theta d\phi ^{2})+6r^{2}d\Omega
_{7}^{2}
\end{equation*}%
and by solving the vacuum Einstein field equations we find: 
\begin{equation*}
F(r)=\frac{3r^{8}+4n^{2}r^{6}+24m}{24r^{6}(r^{2}+n^{2})}
\end{equation*}%
Here $d\Omega _{7}^{2}$ is the metric on the $7$-sphere, normalized such
that its Ricci tensor is $R_{ij}=6g_{ij}$. The Euclidean section is obtained
by analytic continuations $t\rightarrow i\chi $ and $n\rightarrow in$. We
can also replace the sphere element by any other Einstein space of positive
curvature. For example, if we embed the seven dimensional de Sitter solution
we obtain the metric: 
\begin{eqnarray}
ds^{2} &=&\tilde{F}(r)(d\chi +2n\cos \theta d\phi )^{2}+\frac{dr^{2}}{\tilde{%
F}(r)}+(r^{2}-n^{2})(d\theta ^{2}+\sin ^{2}\theta d\phi ^{2})  \notag \\
&&+r^{2}\bigg[-\left( 1-\frac{R^{2}}{6}\right) dt^{2}+\frac{dR^{2}}{1-\frac{%
R^{2}}{6}}+R^{2}d\Omega _{5}^{2}\bigg]  \notag \\
\tilde{F}(r) &=&\frac{3r^{8}-4n^{2}r^{6}+24m}{24r^{6}(r^{2}-n^{2})}
\end{eqnarray}%
It can be easily checked there is a curvature singularity located at $r=0$.
Misner string singularities can be removed by requiring the $\chi $
coordinate to have period $8\pi n$ and $m=\frac{n^{8}}{24}$. The values of
the coordinate $r$ are then restricted to $r>n$ avoiding the curvature
singularity at $r=0$. This solution corresponds to a seven dimensional
fixed-point set of the isometry $\partial _{\chi }$; since this is not the
maximal possible co-dimension, it is a Taub-nut solution.

The other possibility is that of a nine-dimensional fixed-point set of the
isometry $\partial _{\chi }$, located at $r_{b}=2n$. The periodicity of the $%
\chi $ coordinate is still $8\pi n$ but now the values of the $r$ coordinate
are such that $r\geq r_{b}=2n$. This in turn avoids the curvature singularities
located at $r=0$ and $r=n$, provided the value of the mass parameter is $m=-%
\frac{64n^{8}}{3}$.

Let us perform now a Kaluza-Klein compactification along the coordinate $%
\chi $. The reduction ansatz is: 
\begin{equation*}
ds_{11}^{2}=e^{\frac{\varphi }{6}}ds_{10}^{2}+e^{-\frac{4\varphi }{3}}(d\chi
+\mathcal{A})^{2}
\end{equation*}
and we obtain the following ten-dimensional fields: 
\begin{eqnarray}
ds_{10}^{2} &=&\tilde{F}^{\frac{1}{8}}(r)r^{2}\bigg[-\left(1- \frac{R^{2}}{6}%
\right)dt^{2}+\frac{dR^{2}}{1-\frac{R^{2}}{6} } +R^{2}d\Omega _{5}^{2}\bigg]
\notag \\
&&+\tilde{F}^{-\frac{7}{8}}dr^{2}+\tilde{F}^{\frac{1}{8}}(r)(r^{2}-n^{2})(d%
\theta ^{2}+\sin ^{2}\theta d\phi ^{2})  \notag \\
\mathcal{A} &=&2n\cos \theta d\phi,~~~~~~~ e^{-\frac{4\varphi }{3}} =\tilde{F%
}(r)
\end{eqnarray}
Now let us perform a sphere reduction of this solution down to
five-dimensions using the metric ansatz: 
\begin{eqnarray}
ds_{10}^2&=&e^{\sqrt{\frac{5}{12}}\phi}ds_{5}^2+e^{-\sqrt{\frac{3}{20}}%
\phi}d\Omega_5^2
\end{eqnarray}
We obtain the following fields: 
\begin{eqnarray}
ds_{5A}&=&-\tilde{F}^{\frac{1}{3}}r^{\frac{16}{3}}R^{\frac{10}{3}}\left(1-%
\frac{R^2}{6}\right)dt^2+\tilde{F}^{\frac{1}{3}}r^{\frac{16}{3}}R^{\frac{10}{%
3}}\frac{dR^2}{1-\frac{R^2}{6}} +\tilde{F}^{-\frac{2}{3}}r^{\frac{10}{3}}R^{%
\frac{10}{3}}dr^2+\tilde{F}^{\frac{1}{3}}R^{\frac{10}{3}}r^{\frac{10}{3}%
}(r^2-n^2)d\Omega_2^2  \notag \\
\mathcal{A} &=&2n\cos \theta d\phi,~~~~~~~ e^{-\frac{4\varphi }{3}} =\tilde{F%
}(r), ~~~e^{-\sqrt{\frac{3}{20}}\phi}=\tilde{F}^{\frac{5}{24}}r^{\frac{10}{3}%
}R^{\frac{10}{3}}
\end{eqnarray}
which give a solution of the equations of motion derived from the
Lagrangian: 
\begin{equation*}
\mathcal{L}_{5}=eR-\frac{1}{2}e(\partial \varphi )^{2}-\frac{1}{2}%
e(\partial\phi )^{2}-\frac{1}{4}ee^{-\frac{3\varphi }{2}-\sqrt{\frac{5}{12}}%
\phi }(\mathcal{F}_{(2)})^{2}+ee^{\frac{4}{\sqrt{15}}\phi }R_{5}
\end{equation*}
where $R_5$ is the curvature scalar of the unit $5$-sphere. We can dualize $%
\mathcal{F}_{(2)}$ to a $3$-form field strength and we find that the above
solution would describe a non-uniform electric string in five dimensions as
our solution depends explicitly on the fifth dimension $R$. This solution is
very likely to be unstable as in eleven dimensions the `de Sitter horizon'
is delocalised along the noncompact direction $r$.

\section{Conclusions}

\label{conclusion}

In this paper we have constructed higher dimensional Kaluza-Klein brane
solutions with and without cosmological constant, generalizing the original
KK-monopole solution \cite{sorkin,mp}. The simplest generalization in four
dimensions uses the Taub-bolt solution as a seed and the physical interpretation of the final KK-bolt monopole solution was recently clarified in the literature \cite{Edward1,Chamblin1}: it corresponds to a pair of coincident extremal dilatonic black holes with opposite unequal magnetic charges.

In five dimensions we considered cosmological Taub-NUT-like metrics \cite%
{csrm,Page1} and performed similar KK reductions down to four dimensions.
The new feature of these solutions is that the four-dimensional dilaton
acquires a potential term in the Lagrangian as an effect of the cosmological
constant. However their asymptotics are not very appealing physically since
they are not asymptotically flat or asymptotically $(A)dS$. Their metric description simplifies when
considered in the string frame. For our explicit examples the
four-dimensional metric in the string frame is very similar to the $AdS$
form in the $(r, t)$ sector, except for a deficit of solid angle in the
angular sector, which is characteristic for global monopoles. Another
feature of the above constructions is that in five dimensions and in some of
the higher dimensional examples the nut charge and the cosmological constant
are intimately related by a constraint equation imposed by the equations of
motion. This constraint makes it impossible to consider situations in which
either the cosmological constant or the nut charge go to zero. From this
perspective the above monopole solutions are qualitatively distinct from
their predecessors.

On the other hand, in six and seven dimensions we have considered
non-singular Ricci-flat solutions for which one can use the KK trick to
obtain similar KK magnetic brane solutions for which the background spaces
are Ricci flat Bohm spaces of the form $S^{p}\times S^{q}$ and generically
have conical singularities. We considered their further reduction down to
four dimensions on Riemannian spaces of constant curvature and specifically
considered such reductions on spheres. In higher dimensions, we showed that
we can also consider monopole solutions in cosmological backgrounds. All the
above solutions from section \ref{higherdim} have a corresponding extension
for which the cosmological constant is non-zero and are similarly expected
to produce magnetic KK brane solutions upon dimensional reduction. In
contrast with the KK procedure to untwist the $U(1)$-fibration, we have
considered in six dimensions another method, which is known to untwist the
circle fibration, namely Hopf duality in string theory. We extended the Hopf duality
rules to the case of a timelike Hopf-duality of the truncated
six-dimensional Type II theories and applied them to generate charged string
solutions in six-dimensions. By performing sphere reductions we obtained the
corresponding four-dimensional solutions with magnetic charges.

As avenues for further research it would be interesting to use the recently
discovered nut-charged rotating solutions in higher dimensions \cite{Klemm}.
It is known that starting from the four dimensional NUT-charged Kerr
solution, the KK procedure leads to a solution describing a brane/anti-brane
pair carrying opposite unbalanced magnetic charges \cite{Edward1}. The case
of higher dimensional Kerr solution has been explored in \cite{Janssen}
where it was found that for instance in six dimensions one obtains a string
loop instead of a pair of monopoles and anti-monopoles. It would be
interesting to see the effect of the nut charge in these cases. Work on this
is in progress and will be reported elsewhere.

\medskip

{\Large Acknowledgements}

This work was supported by the Natural Sciences \& Engineering Research
Council of Canada.

\renewcommand{\theequation}{A-\arabic{equation}} 
\setcounter{equation}{0} 

\section*{A: Kaluza-Klein Reduction}

\label{KKreduction}

The usual metric ansatz for the dimensional reduction from $(D+1)$ dimensions to $D$
dimensions in Kaluza-Klein theory is given by: 
\begin{eqnarray}
d\hat{s}^2 = e^{2\alpha\varphi}ds^2 + e^{-2(D-2)\alpha\varphi}(dz + \mathcal{%
A})^2  \label{KKansatz}
\end{eqnarray}
with $\alpha=\sqrt{\frac{1}{2(D-1)(D-2)}}$. Here the $D$-dimensional metric $%
ds^2$ corresponds to the so-called Einstein frame, while its conformal
rescaling $e^{2\alpha\varphi}ds^2$ is the metric in the string frame.

Then a general bosonic Lagrangian in $(D+1)$ dimensions of the form: 
\begin{equation*}
\mathcal{L}=\hat{e}\hat{R}+2\hat{e}\lambda -\frac{1}{2}\hat{e}(\partial {%
\hat{\phi}})^{2}-\frac{1}{2n!}\hat{e}e^{\hat{a}\hat{\phi}}\hat{F}_{n}^{2}
\end{equation*}
will reduce in $D$ dimensions to the Lagrangian given by the formula \cite%
{KK}: 
\begin{eqnarray}
\mathcal{L} &=&eR-\frac{1}{2}e(\partial {\phi })^{2}-\frac{1}{2}e(\partial {%
\varphi })^{2}+2e\lambda e^{2\alpha \varphi }-\frac{1}{4}e\cdot
e^{-2(D-1)\alpha \varphi }\mathcal{F}^{2}  \notag \\
&&-\frac{1}{2n!}e\cdot e^{-2(n-1)\alpha \varphi +\hat{a}\phi }{F^{\prime }}%
_{n}^{2}-\frac{1}{2(n-1)!}e\cdot e^{2(D-n)\alpha \varphi +\hat{a}%
\phi}F_{n-1}^{2}  \label{KKlag}
\end{eqnarray}
Here $e=\sqrt{-g}$, $\mathcal{F}=d\mathcal{A}$ and $\mathcal{A}=\mathcal{A}%
_{\mu }dx^{\mu }$ is the one-form potential that appears in the Kaluza-Klein
form of the metric, while the ansatz used for the KK-reduction of the matter
fields from $(D+1)$-dimensions is $\hat{\phi}=\phi $ for the scalar field,
and $\hat{A}_{n-1}~=~A_{n-1}+A_{n-2}\wedge dz$ for the anti-symmetric
potential $\hat{A}_{n-1}$. The field strengths of the potentials $A_{n-1}$
and $A_{n}$ are $F_{n}~=~dA_{n-1}$, respectively $F_{n-1}~=~dA_{n-2}$ and we
have defined $F_{n}^{\prime }=F_{n}-F_{n-1}\wedge \mathcal{A}$.

Note that the presence of the cosmological constant in the higher
dimensional theory induces a scalar potential for the Kaluza-Klein scalar
field $\varphi $. Also, if the isometry generated by the Killing vector $%
\frac{\partial}{\partial z}$ has fixed points, then the dilaton $\varphi$
will diverge and the $D$-dimensional metric will be singular at those points.

\renewcommand{\theequation}{B-\arabic{equation}} 
\setcounter{equation}{0} 

\section*{B: T-duality in six dimensions}

The Lagrangian in $D=6$ obtained by dimensional reduction of Type IIB on a
torus and after performing a consistent truncation is given by \cite{Duff2}: 
\begin{eqnarray}
\mathcal{L}_{6B}&=&eR-\frac{1}{2}e(\partial\phi_1)^2-\frac{1}{2}%
e(\partial\phi_2)^2-\frac{1}{2}ee^{2\phi_1}(\partial\chi_1)^2-\frac{1}{2}%
ee^{2\phi_2}(\partial\chi_2)^2  \notag \\
&& -\frac{1}{12}ee^{-\phi_1-\phi_2}(F^{NS}_{(3)})^2-\frac{1}{12}%
ee^{\phi_1-\phi_2}(F^{RR}_{(3)})^2+\chi_2dA^{NS}_{(2)}\wedge dA^{RR}_{(2)}
\label{6IIB}
\end{eqnarray}
where $F_{(3)}^{NS}=dA^{NS}_{(2)}$ and $F^{RR}_{(3)}=dA^{RR}_{(2)}+%
\chi_1dA_{(2)}^{NS}$. This Lagrangian is related by T-duality in $D=5$ to a
different six-dimensional theory obtained by making a consistent truncation
of Type IIA compactified on a four-dimensional torus. The corresponding
Lagrangian is given by: 
\begin{eqnarray}
\mathcal{L}_{6A}&=&eR-\frac{1}{2}e(\partial\phi_1)^2-\frac{1}{2}e(\partial\phi_2)^2 -\frac{1}{48}ee^{\frac{
\phi_1}{2}-\frac{3\phi_2}{2}}(F_{(4)})^2 -\frac{1}{12}ee^{-\phi_1-%
\phi_2}(F_{(3)})^2\nonumber\\ 
&&-\frac{1}{4}ee^{\frac{3\phi_1}{2}-\frac{\phi_2}{2}}(F_{(2)})^2  \label{6IIA}
\end{eqnarray}
where $F_{(4)}=dA_{(3)}-dA_{(2)}\wedge A_{(1)}$, $F_{(3)}=dA_{(2)}$
corresponds to the NS-NS 3-form $F_{(3)1}$ and $F_{(2)}=dA_{(1)}$ is the RR
2-form $\mathcal{F}^1_{(2)}$, with the index `1' denoting here the first
reduction step from $D=11$ to $D=10$.

Let us focus on Type IIA theory first. Under a dimensional reduction using
the formulae from the previous appendix we have: 
\begin{eqnarray}
ds_{6}^{2}&=&e^{\frac{\varphi}{\sqrt{6}}}ds_{5}^2+e^{\frac{-3\varphi}{\sqrt{6%
}}}(dz+\mathcal{A}_{(1)})^2
\end{eqnarray}
and we obtain the following 5-dimensional Lagrangian: 
\begin{eqnarray}
\mathcal{L}_{5A}&=&eR-\frac{1}{2}e(\partial\phi_1)^2-\frac{1}{2}%
e(\partial\phi_2)^2-\frac{1}{2}e(\partial\varphi)^2-\frac{1}{48}ee^{-\frac{%
3\varphi}{\sqrt{6}}+\frac{\phi_1}{2}-\frac{3\phi_2}{2}}(F_{(4)}^{\prime})^2 
\notag \\
&&-\frac{1}{12}ee^{\frac{\varphi}{\sqrt{6}}+\frac{\phi_1}{2}-\frac{3\phi_2}{2%
}}(F_{(3)1})^2-\frac{1}{12}ee^{-\frac{2\varphi}{\sqrt{6}}-\phi_1-%
\phi_2}(F_{(3)}^{\prime})^2-\frac{1}{2}ee^{-\frac{4\varphi}{\sqrt{6}}}%
\mathcal{F}_{(2)}^2  \notag \\
&&-\frac{1}{4}ee^{-\frac{\varphi}{\sqrt{6}}+\frac{3\phi_1}{2}-\frac{\phi_2}{2%
}}(F_{(2)}^{\prime})^2-\frac{1}{4}ee^{\frac{2\varphi}{\sqrt{6}}%
-\phi_1-\phi_2}(F_{(2)1})^2-\frac{1}{2}ee^{\frac{3\varphi}{\sqrt{6}}+\frac{%
3\phi_1}{2}-\frac{\phi_2}{2}}(d A_{(0)1})^2  \label{5A}
\end{eqnarray}
where the field strengths are defined as follows (see the formulae from
appendix A): 
\begin{eqnarray}
F_{(2)}^{\prime}&=&dA_{(1)}-dA_{(0)1}\wedge\mathcal{A}_{(0)},
~~~~~~~F_{(3)}^{\prime}=dA_{(2)}-dA_{(1)}\wedge\mathcal{A}_{(1)}  \notag \\
F_{(3)1}&=&dA_{(2)1}+dA_{(1)}\wedge A_{(1)}-dA_{(2)}\wedge A_{(0)1}, ~~~~~
F_{(4)}^{\prime}=dA_{(3)}-dA_{(2)}\wedge A_{(1)}-F_{(3)1}\wedge\mathcal{A}%
_{(1)}  \notag
\end{eqnarray}
while $\mathcal{F}_{(2)}=d\mathcal{A}_{(1)}$ and $F_{(2)1}=dA_{(1)1}$. Upon
dualising $F_{(4)}$ to a 1-form field strength $d\chi^{\prime}$ its kinetic
term in the above Lagrangian will be replaced by: 
\begin{eqnarray}
-\frac{1}{2}ee^{\frac{3\varphi}{\sqrt{6}}-\frac{\phi_1}{2}+\frac{3\phi_2}{2}%
}(d\chi^{\prime})^2+\chi^{\prime}F_{(3)}^{\prime}\wedge
F_{(2)}^{\prime}+\chi^{\prime}F_{(3)1}\wedge\mathcal{F}_{(2)}
\end{eqnarray}
If we perform the field redefinitions field redefinitions: 
\begin{eqnarray}
A_{(1)}^{\prime}&=&A_{(1)}-A_{(0)1}\wedge\mathcal{A}_{(1)}, ~
A_{(2)}^{\prime}=A_{(2)}-A_{(1)1}\wedge\mathcal{A}_{(1)}, ~~
A_{(2)1}^{\prime}=A_{(2)1}+A_{(1)1}\wedge A_{(1)}^{\prime}
\end{eqnarray}
we find: 
\begin{eqnarray}
F_{(2)}^{\prime}=dA_{(1)}^{\prime}+A_{(0)1}\wedge \mathcal{F}_{(2)}, ~~~~~~~
F_{(3)}^{\prime}=dA_{(2)}^{\prime}-A_{(1)1}\wedge\mathcal{F}_{(2)}  \notag \\
F_{(3)1}=dA_{(2)1}^{\prime}+dA_{(1)}^{\prime}\wedge
A_{(1)1}-A_{(0)1}(dA_{(2)}^{\prime}-A_{(1)1}\wedge\mathcal{F}_{(2)})  \notag
\\
\chi^{\prime}F_{(3)}^{\prime}\wedge
F_{(2)}^{\prime}+\chi^{\prime}F_{(3)1}\wedge\mathcal{F}_{(2)}=\chi^{%
\prime}(dA_{(2)}^{\prime}\wedge dA_{(1)}^{\prime}+dA_{(2)1}^{\prime}\wedge%
\mathcal{F}_{(2)})
\end{eqnarray}
Similarly, for the dimensional reduction of Type IIB Lagrangian we obtain: 
\begin{eqnarray}
\mathcal{L}_{5B}&=&eR-\frac{1}{2}e(\partial\phi_1)^2-\frac{1}{2}%
e(\partial\phi_2)^2-\frac{1}{2}e(\partial\varphi)^2-\frac{1}{2}%
ee^{2\phi_1}(\partial\chi_1)^2-\frac{1}{2}ee^{2\phi_2}(\partial\chi_2)^2 
\notag \\
&&-\frac{1}{12}ee^{-\frac{2\varphi}{\sqrt{6}}+\phi_1-\phi_2}(F_{(3)}^{^{%
\prime}~RR})^2-\frac{1}{12}ee^{-\frac{2\varphi}{\sqrt{6}}-\phi_1-%
\phi_2}(F_{(3)}^{^{\prime}~NS})^2-\frac{1}{2}ee^{-\frac{4\varphi}{\sqrt{6}}}%
\mathcal{F}_{(2)}^2-\frac{1}{4}ee^{\frac{2\varphi}{\sqrt{6}}%
+\phi_1-\phi_2}(F_{(2)1}^{RR})^2  \notag \\
&&-\frac{1}{4}ee^{\frac{2\varphi}{\sqrt{6}}-\phi_1-\phi_2}(F_{(2)1}^{NS})^2-%
\chi_2dA_{(2)}^{RR}\wedge dA_{(1)1}^{NS}+\chi_2dA_{(2)}^{NS}\wedge
dA_{(1)1}^{RR}  \label{5B}
\end{eqnarray}
where $F_{(2)1}^{NS}=dA_{(1)1}^{NS}$, $\mathcal{F}_{(2)}=d\mathcal{A}_{(1)}$
and: 
\begin{eqnarray}
F_{(3)}^{^{\prime}~NS}=dA_{(2)}^{NS}-dA_{(1)1}^{NS}\wedge\mathcal{A}_{(1)},
~~~~~~~F_{(2)1}^{RR}=dA_{(1)1}^{RR}+\chi_1 dA_{(1)1}^{NS}  \notag \\
F_{(3)}^{^{\prime}~RR}=dA_{(2)}^{RR}-dA_{(1)1}^{RR}\wedge \mathcal{A}%
_{(1)}+\chi_1dA_{(2)}^{NS}\wedge\mathcal{A}_{(1)}
\end{eqnarray}
As shown first in \cite{Duff2}, the $T$-duality rules relating the two
truncated theories (\ref{5A}) and (\ref{5B}) are: 
\begin{eqnarray}
A_{(0)1}\rightarrow\chi_1, ~~~A_{(1)1}\rightarrow\mathcal{A}_{(1)}, ~~~%
\mathcal{A}_{(1)}\rightarrow A_{(1)1}^{NS},~~~\chi^{\prime}\rightarrow\chi_2,
\notag \\
A_{(1)}^{\prime}\rightarrow A_{(1)1}^{RR}, ~~~ A_{(2)}^{\prime}\rightarrow
A_{(2)}^{NS}, ~~~A_{(2)1}^{\prime}\rightarrow - A_{(2)}^{RR}
\end{eqnarray}
together with a rotation of the scalars: 
\begin{eqnarray}
\left(%
\begin{array}{c}
\phi_1 \\ 
\phi_2 \\ 
\varphi
\end{array}%
\right)_{II A,B}&=&\left(%
\begin{array}{ccc}
\frac{3}{4} & -\frac{1}{4} & -\frac{\sqrt{6}}{4} \\ 
-\frac{1}{4} & \frac{3}{4} & -\frac{\sqrt{6}}{4} \\ 
-\frac{\sqrt{6}}{4} & -\frac{\sqrt{6}}{4} & -\frac{1}{2}%
\end{array}%
\right)\left(%
\begin{array}{c}
\phi_1 \\ 
\phi_2 \\ 
\varphi
\end{array}%
\right)_{II B,A}
\end{eqnarray}
which takes care of the dilaton couplings of the field strengths.

We are also interested in performing a timelike $T$-duality. As it is known
this duality will relate Type IIA (respectively IIB) to Type IIB$^{\ast }$
(respectively IIA$^{\ast }$). We wish to see if at the level of our
truncated theories the timelike T-duality rules are still valid.

Consider first the Type IIA theory. Upon a timelike dimensional reduction
the Lagrangian of the reduced theory will have a form similar with (\ref{5A}%
); however the kinetic terms for $F_{(3)1}$, $F_{(2)1}$, $\mathcal{F}_{(2)}$
and $dA_{(0)1}$ will have the reversed sign \cite{Cremmer}. The
five-dimensional metric is now of Euclidean signature and when we dualize
the $4$-form $F_{(4)}^{\prime }$ to a scalar field strength $\chi ^{\prime }$
we obtain a positive kinetic term for this scalar. We expect to be able to
relate this theory to a timelike reduction of a truncated six-dimensional
Type IIB$^{\ast }$ theory by applying the $T$-duality rules given above. Now, it is known that the action of Type IIB$^{\ast }$ in ten dimensions is obtained from the usual Type IIB action after we reverse the
signs of the $RR$ kinetic terms. As the sign of such kinetic terms was
irrelevant when discussing the truncation to six dimensions we see that a
consistent truncation of Type IIB$^{\ast }$ in six dimensions will be given
by the Lagrangian (\ref{6IIB}) in which we must reverse the sign on the
kinetic terms for the $RR$ fields, i.e. we must reverse the sign of the
kinetic terms for $F_{(3)}^{RR}$ and also for $\chi _{2}$ (which appears
from the dualisation of the $RR$ field $B_{(4)}$). When performing a
timelike dimensional reduction the final Type IIB$^{\ast }$ Lagrangian will
be similar with (\ref{5B}) with reverted signs for the kinetic terms of $%
\chi _{2}$, $F_{(3)}^{RR}$, $F_{(2)1}^{NS}$ respectively $\mathcal{F}_{(2)}$%
. It is then straightforward to see that the $T$-duality will relate our
truncated Type IIA theory with the truncated Type IIB$^{\ast }$. It is easy
to extend the above considerations to show that a timelike $T$-duality will
relate Type IIB with Type IIA$^{\ast }$ at the level of our truncated
theories. Also Type IIA$^{\ast }$ and Type IIB$^{\ast }$ are related by a
usual $T$-duality along a spacelike direction.


\begin{thebibliography}{99}
\bibitem{sorkin} Rafael~D.~Sorkin ``Kaluza-Klein Monopole'' Phys.\ Rev.\
Lett.\ B\ \textbf{51} (1983) 87.

\bibitem{mp} D.~J.~Gross and M.~J.~Perry ``Magnetic Monopoles in
Kaluza-Klein theory'' Nucl.\ Phys.\ \textbf{B226} (1983) 29.

\bibitem{onemli} V.~K.~Onemli and B.~Tekin, ``Kaluza-Klein monopole in AdS
spacetime,'' [arXiv:hep-th/0301027]. 

\bibitem{d2} D.~Astefanesei, R.~B.~Mann and E.~Radu, ``Nut charged
space-times and closed timelike curves on the boundary,''
[arXiv:hep-th/0407110]
;~D.~Astefanesei and E.~Radu ``Quantum effects in a rotating
spacetime'', Int.\ J.\ Mod.\ Phys.\ D \textbf{11} 715 (2002)
[arXiv:gr-qc/0112029].

\bibitem{csrm} R.~Mann and C.~Stelea,``Nuttier (A)dS black holes in higher dimensions,''
  Class.\ Quant.\ Grav.\  {\bf 21}, 2937 (2004),
  [arXiv:hep-th/0312285].

\bibitem{Page1} H.~Lu, D.~N.~Page and C.~N.~Pope, ``New inhomogeneous
Einstein metrics on sphere bundles over Einstein-Kaehler manifolds,''
[arXiv:hep-th/0403079]. 

\bibitem{rick} R.~Clarkson,~R.~B.~Mann, ``Eguchi-Hanson Solitons'', [arXiv:hep-th/00508109];~
R.~Clarkson,~R.~B.~Mann, ``Eguchi-Hanson Solitons in Odd Dimensions'' , [arXiv:hep-th/0508200].

\bibitem{Misner} C.~W.~Misner ``The Flatter Regions of Newman, Unti and
Tamburino's Generalized Schwarzschid Space'' J.\ Math.\ Phys. \textbf{4}
(1963) 924.

\bibitem{Page} D.~N.~Page ``Taub-NUT instanton with a Horizon'' Phys.\
Lett.\ B\ \textbf{78} (1978) 249-251.

\bibitem{Mann} R.B. Mann, ``Misner String\ Entropy'', Phys. Rev. D \textbf{60%
}, 104047 (1999) [arXiv:hep-th/9903229].

\bibitem{Chamblin} A.~Chamblin, R.~Emparan, C.~V.~Johnson and R.~C.~Myers,
``Large N phases, gravitational instantons and the NUTs and bolts of AdS
holography,'' Phys.\ Rev.\ D \textbf{59}, 064010 (1999)
[arXiv:hep-th/9808177].

\bibitem{Edward1} Y.~C.~Liang and E.~Teo, ``Black diholes with unbalanced
magnetic charges,'' Phys.\ Rev.\ D \textbf{64}, 024019 (2001)
[arXiv:hep-th/0101221]. 

\bibitem{Chamblin1} A.~Chamblin and R.~Emparan, ``Bubbles in Kaluza-Klein
theories with space- or time-like internal dimensions,'' Phys.\ Rev.\ D 
\textbf{55}, 754 (1997) [arXiv:hep-th/9607236]. 

\bibitem{Demiansky} M.~Demiansky and E.~T.~Newman, ``A combined Kerr-NUT
solution of the Einstein field equations'' Bull.\ Acad.\ Polon.\ Sci.\ Ser.\
Math.\ Astron.\ Phys.\ \textbf{14} 653(1966).

\bibitem{Gibbons:nf} G.~W.~Gibbons and M.~J.~Perry, ``New Gravitational
Instantons And Their Interactions,'' Phys.\ Rev.\ D \textbf{22}, 313 (1980). 

\bibitem{vilenkin} M.~Barriola and A.~Vilenkin, ``Gravitational Field Of A
Global Monopole,'' Phys.\ Rev.\ Lett.\ \textbf{63}, 341 (1989). 

\bibitem{sparks}
  J.~F.~Sparks,``Kaluza-Klein branes,''
  [arXiv:hep-th/0105209].

\bibitem{awad}
  A.~Awad and A.~Chamblin,
  ``A bestiary of higher dimensional Taub-NUT-AdS spacetimes,''
  Class.\ Quant.\ Grav.\  {\bf 19}, 2051 (2002)
  [arXiv:hep-th/0012240].

\bibitem{bais} F.~A.~Bais and P.~Batenberg, ``A New Class of
Higher-Dimensional Kaluza-Klein Monopole and Instanton Solutions'', Nucl.\
Phys.\ B \textbf{253}, 162 (1985).

\bibitem{Bohm} C.~Bohm,``Inhomogenous Einstein metrics on low-dimensional
spheres and other low-dimensional spaces,'' Invent.\ Math.\ \textbf{134},
145 (1998).

\bibitem{Gibbons}  G.~W.~Gibbons, S.~A.~Hartnoll and C.~N.~Pope,  ``Bohm and
Einstein-Sasaki metrics, black holes and cosmological event  horizons,'' 
Phys.\ Rev.\ D \textbf{67}, 084024 (2003)  [arXiv:hep-th/0208031]. 

\bibitem{bremer} M.~S.~Bremer, M.~J.~Duff, H.~Lu, C.~N.~Pope and
K.~S.~Stelle, ``Instanton cosmology and domain walls from M-theory and
string theory,'' Nucl.\ Phys.\ B \textbf{543}, 321 (1999)
[arXiv:hep-th/9807051].

\bibitem{hartnoll}
  S.~A.~Hartnoll,``Axisymmetric non-abelian BPS monopoles from G(2) metrics,''
  Nucl.\ Phys.\ B {\bf 631}, 325 (2002)
  [arXiv:hep-th/0112235].

\bibitem{Duff1} M.~J.~Duff, H.~Lu and C.~N.~Pope, ``AdS(5) x S(5)
untwisted,'' Nucl.\ Phys.\ B \textbf{532}, 181 (1998)
[arXiv:hep-th/9803061]. 

\bibitem{Duff2} M.~J.~Duff, H.~Lu and C.~N.~Pope, ``AdS(3) x S**3
(un)twisted and squashed, and an O(2,2,Z) multiplet of dyonic strings,''
Nucl.\ Phys.\ B \textbf{544}, 145 (1999) [arXiv:hep-th/9807173]. 

\bibitem{Cvetic} M.~Cvetic, H.~Lu and C.~N.~Pope, ``Consistent warped-space
Kaluza-Klein reductions, half-maximal gauged supergravities and CP(n)
constructions,'' Nucl.\ Phys.\ B \textbf{597}, 172 (2001)
[arXiv:hep-th/0007109]. 

\bibitem{d1}
  D.~Astefanesei and G.~C.~Jones,
  ``S-branes and (anti-)bubbles in (A)dS space'',
  [arXiv:hep-th/0502162];~ D.~Astefanesei, R.~B.~Mann and C.~Stelea,``Nuttier bubbles,''
  [arXiv:hep-th/0508162].
  
\bibitem{Hull}  C.~M.~Hull,  ``Timelike T-duality, de Sitter space, large N
gauge theories and  topological field theory,''  JHEP \textbf{9807}, 021
(1998)  [arXiv:hep-th/9806146]. 

\bibitem{KK} H.~Lu, C.~N.~Pope, E.~Sezgin and K.~S.~Stelle, ``Stainless
super p-branes,'' Nucl.\ Phys.\ B \textbf{456}, 669 (1995)
[arXiv:hep-th/9508042].

\bibitem{Cremmer}  E.~Cremmer, I.~V.~Lavrinenko, H.~Lu, C.~N.~Pope,
K.~S.~Stelle and T.~A.~Tran,  ``Euclidean-signature supergravities,
dualities and instantons,''  Nucl.\ Phys.\ B \textbf{534}, 40 (1998) 
[arXiv:hep-th/9803259]. 

\bibitem{Klemm} D.~Klemm, ``Rotating black branes wrapped on Einstein
spaces,'' JHEP \textbf{9811}, 019 (1998) [arXiv:hep-th/9811126]. 

\bibitem{Janssen} B.~Janssen and S.~Mukherji, ``Kaluza-Klein dipoles,
brane/anti-brane pairs and instabilities,'' [arXiv:hep-th/9905153]. 





\end{thebibliography}
\end{document}